\documentclass[aps,prd,showpacs,amsmath,amssymb]{revtex4}
\usepackage{epsfig}

\begin{document}

\title{ An additional symmetry in the Weinberg -- Salam model.}
{
\vspace{1cm}

\author{\firstname{B.L.G.}~\surname{Bakker}}
\affiliation{\rm Department of Physics and Astronomy, Vrije Universiteit,
Amsterdam, The Netherlands.}
\author{\firstname{A.I.}~\surname{Veselov}}
\affiliation{\rm ITEP, B.Cheremushkinskaya 25, Moscow, 117259, Russia.}
\author{\firstname{M.A.}~\surname{Zubkov}}
\affiliation{\rm ITEP, B.Cheremushkinskaya 25, Moscow, 117259, Russia.}


\begin{abstract}
An additional $Z_6$ symmetry hidden in the fermion and Higgs sectors of the
Standard Model  has been found recently. It has a singular nature and is
connected to the centers of the $SU(3)$ and $SU(2)$ subgroups of the gauge
group. A lattice regularization of the Standard Model was constructed that
possesses this symmetry. In this paper we report our results on the numerical
simulation of
 its Electroweak sector.
\end{abstract}


\maketitle

\section{Introduction}

It is the conventional point of view that all the symmetries of the Standard
Model (SM), which must be used when dealing with its discretization, are known.
Recently it was found that there exists an additional $Z_6 = Z_2 \otimes Z_3$
symmetry in the fermion and Higgs sectors of the SM \cite{BVZ2003}. It has a
singular nature and is connected to the centers $Z_3$ and $Z_2$ of the $SU(3)$
and $SU(2)$ subgroups\footnote{The emergence of $Z_6$ symmetry in the Standard
Model and its supersymmetric extension was considered in different context in
\cite{Z6}}. The gauge sector of the SM (in its discretized form) was redefined
in such a way that it has the same perturbation expansion as the original one,
while keeping the mentioned symmetry.  The resulting model differs from the
conventional SM via its symmetry properties. Therefore, we expect it would
describe nature better than the  conventional discretized Standard Model, if
the additional symmetry does takes place.

It is worth mentioning that the present status of the Standard Model on the
lattice implies that it must be considered as a finite cutoff theory
\cite{SM_present_status}. This is in agreement with the understanding that the
SM does not describe physics at extremely small distances. Hence, it is
sufficient to consider a cutoff $\Lambda$ that is finite but much larger than
all observed energies. The consideration of the infinite cutoff limit would be
an attempt to continue the Standard Model to infinitesimal distances. Now it is
believed that this attempt leads to a trivial continuum theory
\cite{triviality}. Nevertheless, at energies much less than the cutoff we can
calculate any physical variable.

So we can examine our model in order to understand whether the considered
additional symmetry is important for the discretization of the Standard Model
or not. As a first step in this direction we investigate numerically the
quenched Electroweak sector  of the constructed discretized SM.

\section{A hidden symmetry}

In this section we repeat our construction reported in \cite{BVZ2003} in
continuum notation. This is done in order to demonstrate the universal
(regularization - independent) nature of the additional symmetry found.

\subsection{The Standard Model}

The Standard Model contains the following variables:

1. The gauge fields associated with the symmetry group $SU(3) \times SU(2)
\times U(1)$, which are the elements of the corresponding algebras:
\begin{eqnarray}
Z_i & = & Z_i^a \lambda_a \in su(3), \nonumber \\
A_i & = & A_i^b \sigma_b \in su(2), \nonumber \\
B_i \in u(1) & = & (-\infty, \infty).
\end{eqnarray}
(Here, $\lambda_a$ are the Gell-Mann matrices, and $\sigma_b$ are the Pauli
matrices.) The corresponding $SU(3)$, $SU(2)$, and $U(1)$ field strengths are:
\begin{eqnarray}
H_{ij} & = & \partial_{[i}Z_{j]} + i [Z_i,Z_j], \nonumber\\
G_{ij} & = & \partial_{[i}A_{j]} + i [A_i,A_j], \nonumber\\
F_{ij} & = & \partial_{[i}B_{j]}. \label{ZGF}
\end{eqnarray}

2. Anticommuting spinor variables, representing leptons and quarks:
\begin{equation}
 \left(
 \begin{array}{ccc}
 e & \mu & \tau \\
 \nu_e & \nu_{\mu} &  \nu_{\tau},
 \end{array}
 \right) \, , \quad
 \left(
 \begin{array}{ccc}
 u & c & t \\
 d & s & b
 \end{array}
 \right) \, .
\end{equation}

3. A scalar doublet
\begin{equation}
\Phi^{\alpha}, \;\alpha = 1,2.
\end{equation}

The action has the form:
\begin{equation}
 S = S_g + S_H + S_f,
\end{equation}
where we denote the fermion part of the action by $S_{f}$, the pure gauge part
is denoted by $S_g$, and the scalar part of the action by $S_H$.

As usual, we consider  $S_g$ in the form:
\begin{equation}
 S_g  =  \frac{1}{4} \int d^4 x \left[\frac{1}{3g_{SU(3)}^2} {\rm Tr}\, H^2
 + \frac{1}{2g_{SU(2)}^2} {\rm Tr}\, G^2 \right. 
  \left. + \frac{1}{g_{U(1)}^2} F^2 \right],
\end{equation}
where we introduced the gauge couplings $g_{SU(3)}$, $g_{SU(2)}$, and $g_{U(1)}
$.

The scalar part of the action is
\begin{equation}
 S_H = \int d^4x\, |(\partial_{\mu} + i A_{\mu} + i B_{\mu})\Phi|^2
     + \int d^4x\, V(|\Phi|),
\end{equation}
where $V(|\Phi|)$ is the potential, which has a minimum at a nonzero value of
$\Phi = v$, causing spontaneous symmetry breaking.

We express $S_f$ through left-handed doublets $L$ and right-handed singlets $R$
of fermions:
\begin{equation}
\begin{array}{ccc}
 L^{\ell}_1 = \frac{1 - \gamma_5}{2}
 \left(\begin{array}{c} e \\ \nu_e \end{array} \right)\, , &
 L^{\ell}_2 = \large{\frac{1 - \gamma_5}{2}}
 \left(\begin{array}{c} \mu \\ \nu_{\mu} \end{array} \right)\, , &
 L^{\ell}_3 = \frac{1 - \gamma_5}{2}
 \left(\begin{array}{c} \tau \\ \nu_{\tau} \end{array} \right)\, , \\
 L^q_1 = \frac{1 - \gamma_5}{2}
 \left(\begin{array}{c} u \\ d \end{array} \right)\, , &
 L^q_2 = \frac{1 - \gamma_5}{2}
 \left(\begin{array}{c} c \\ s \end{array} \right)\, , &
 L^q_3 = \frac{1 - \gamma_5}{2}
 \left(\begin{array}{c} t \\ b \end{array} \right)\, , \\
 R^{\ell}_1 = \frac{1 + \gamma_5}{2} e, &
 R^{\ell}_2 = \frac{1 + \gamma_5}{2} \mu, &
 R^{\ell}_3 = \frac{1 + \gamma_5}{2} \tau, \\
 R^q_{1,1} = \frac{1 + \gamma_5}{2} u, &
 R^q_{1,2} = \frac{1 + \gamma_5}{2} c, &
 R^q_{1,3} = \frac{1 + \gamma_5}{2} t,  \\
 R^q_{2,1} = \frac{1 + \gamma_5}{2} d, &
 R^q_{2,2} = \frac{1 + \gamma_5}{2} s, &
 R^q_{2,3} = \frac{1 + \gamma_5}{2} b.
\end{array}
\end{equation}
The fermion part of the action is:
\begin{equation}
 S_f  =  \int d^4 x\, \{{\cal L}^{\rm L}_{\ell}+{\cal L}^{\rm R}_{\ell}
 + {\cal L}^{\rm L}_{\rm q}
 + {\cal L}^{{\rm R},1}_{\rm q}  + {\cal L}^{{\rm R},2}_{\rm q} 
 +  {\cal L}^{\ell}_{\rm mass} +  {\cal L}^q_{\rm mass}\}.
\end{equation}
Here,
\begin{eqnarray}
 {\cal L}^{\rm L}_{\ell} & = & i \bar{L}^{\ell}_i
 (\partial_{\mu} + i A_{\mu} - i B_{\mu}) \gamma_{\mu}L^{\ell}_i
 ,\nonumber \\
 {\cal L}^{\rm R}_{\ell} & = & i \bar{R}^{\ell}_i
 (\partial_{\mu} - 2 i B_{\mu})\gamma_{\mu} R^{\ell}_i, \nonumber \\
 {\cal L}^{\rm L}_{\rm q} & = & i \bar{L}^q_i
 (\partial_{\mu}  + i Z_{\mu} +  i A_{\mu} + (i/3) B_{\mu})\gamma_{\mu} L^q_i
, \nonumber \\
 {\cal L}^{{\rm R},1}_{\rm q} & = & i \bar{R}^q_{1,i}
 (\partial_{\mu}  + i Z_{\mu} + (4i/3) B_{\mu}) \gamma_{\mu}R^q_{1,i}
,\nonumber \\
 {\cal L}^{{\rm R},2}_{\rm q} & = & i \bar{R}^q_{2,i}
 (\partial_{\mu}  + i Z_{\mu}  - (2i/3) B_{\mu}) \gamma_{\mu} R^q_{2,i}
,\nonumber \\
 {\cal L}^q_{\rm mass} & = & \frac{1}{v} \sum_{i} m^{q}_i
 (\bar{L}^q_i)^{\alpha} \Phi^{\alpha} R^q_{1,i} + \frac{1}{v} \sum_{ij} M_{ij}
 (\bar{L}^q_i)^{\alpha} \Omega^{\alpha} R^q_{2,j} + {\rm h.c.},\nonumber\\
 {\cal L}^{\ell}_{\rm mass} & = & \frac{1}{v}\sum_i m^{\ell}_i
 (\bar{L}^{\ell}_i)^{\alpha} \Phi^{\alpha} R^{\ell}_i + {\rm h.c.}.
\label{action_f}
\end{eqnarray}
In these expressions $\Omega = i\sigma_2 \Phi$ ($i\sigma_2$ is the charge
conjugation operator), $\bar{\psi} = \psi^\dag \gamma_0$, and
\begin{eqnarray}
 m^{\ell}_1 & = & m_e, \, m^{\ell}_2 = m_{\mu}, \, m^{\ell}_3 = m_{\tau}, \nonumber\\
 m^q_1 & = & m_u,\, m^q_2 = m_c,\, m^q_3 = m_t.
\end{eqnarray}
$M$ is the mass matrix, which eigenvalues represent the masses of the $d$, $s$,
and $b$ quarks. The nondiagonality of this matrix gives rise to the phenomenon
of quark mixing.

All necessary information about the Euclidian dynamics of the SM is contained
in the gauge invariant correlators:
\begin{equation}
 \langle O({\rm fields}) \rangle  =   \int D Z\,  D A \, D B \,D e
 \,D \bar{e}\, D \nu_e D \bar{\nu_e}\, \dots \,D \Phi \, 
  \exp( - S({\rm fields} ) ) \, O({\rm fields})
\label{cor}
\end{equation}

\subsection{Representation of the Standard Model in loop space}

The hidden symmetry we are talking about may be seen after reformulation of the
Standard Model through loop variables. (For the definition of the notations
connected with loop space dynamics see \cite{loop_equations}.) The derivation
is as follows.

First, we note that in Eq.~(\ref{action_f}) $L$ and $R$ can be treated as
independent two-component Weil spinors. In the Weil basis of the $\gamma$ -
matrices the Euclidean fermion Lagrangian contains quadratic terms like $ \,
L^\dagger (\nabla_0 - i \nabla_{i} \sigma_{i}) L$ and $\,R^\dagger (\nabla_0 +
i \nabla_{i} \sigma_{i}) R $ (where $\nabla_{\mu}$ is a covariant derivative,
$\sigma_i$ ($i = 1,2,3$) are Pauli matrices), and  an interaction term like
$(L^\dagger \Phi) R$.

For an arbitrary gauge invariant correlator we have:
\begin{equation}
 \langle O({\rm fields}) \rangle  =  \int D Z\,  D A \, D B\,
 \exp ( - S_g) \langle O({\rm fields}) \rangle_{{ f}, \, { s}} .
\end{equation}
Here, $ \langle O({\rm fields}) \rangle_{{ f}, \, { s}}$ is an integral over
fermions and over the scalar field. It is calculated in the model with an
external gauge field. First we perform an integration over the Grassmann
variables. We can do it using simple Feynman rules. The diagrams contain
propagators of Weil spinors, correlation functions of the scalar field, and
interaction vertices that are coming from the term $L^{\dagger\, \alpha}
\Phi^{\alpha} R$. The  loops coming from the fermion determinant should also be
taken into account. We use the path integral representation of the propagators
and of the fermion determinant (see, for example, \cite{loop_equations,
path_integral}). In order to calculate the scalar field correlators we use the
lattice regularization.  It is well - known that the bosonic path integral for
any field correlators in the external gauge field on the lattice has a
representation as a sum over all possible closed loops. (For the details of the
calculation see, for example, \cite{lattice_paths}).
 After returning to the continuum representation we arrive at a
path-integral representation of the scalar field correlators.

Finally we represent any correlator Eq.~(\ref{cor}) in the form:
\begin{eqnarray}
 \langle O({\rm fields}) \rangle & = &
 \!\! \int D {\cal C}_{\alpha} {\cal O}({\cal C})  \int D Z\,  D A\,  D B \,
 \exp( - S_g ) {\cal W} ({\cal C}) = \nonumber\\
 & = &  \!\!\int D {\cal C}_{\alpha} {\cal O}({\cal C})
 \langle {\cal W} ({\cal C})\rangle \label{cor1}.
\end{eqnarray}

Here, ${\cal C}_{\alpha}$ stands for the set of paths.  Each path corresponds
to one of the fermions (left - or right - handed) or to the scalar. It may
either be closed or ending in a vertex. Each vertex corresponds to the
transformation of left-handed fermion into right-handed ones and emission or
absorption of a scalar. The definition of the measure $D {\cal C}_{\alpha}$
comes from the path integral representations of the bosonic correlator
mentioned above, the fermion determinant, and the fermion propagator. It
includes all possible paths described above.  The index $\alpha$ enumerates all
fermions and the scalar. The functionals $\cal O$ do not depend upon the gauge
fields and are, hence, not of interest to us. The full dependence on the gauge
fields is now concentrated in the loop variable $\cal W$ that is simply a
product of parallel transporters $W({\cal C}_{\alpha})$ corresponding to the
fermions and to the scalar:
\begin{equation}
{\cal W} ({\cal C}) = {\rm Tr}\, \Pi  W({\cal C}_{\alpha}). \label{WWW}
\end{equation}

In Eq.~(\ref{WWW}) we encounter six different parallel transporters:
\begin{eqnarray}
 W_{{\rm L}_{\ell}}({\cal C}) & = & \, P \, {\rm exp}( i \int_{\cal C}
 ( A_{\mu} -  B_{\mu}) d x_{\mu}) = \omega_{SU(2)} \omega^{-1}_{U(1)}, \nonumber\\
 W_{{\rm R}_{\ell}}({\cal C}) & = & {\rm exp}( i \int_{\cal C}
 (-2  B_{\mu}) d x_{\mu}) = \omega_{U(1)}^{-2}, \nonumber\\
 W_{{\rm L}_{q}}({\cal C}) & = & \, P \, {\rm exp}( i \int_{\cal C}
 (Z_{\mu} + A_{\mu} + \mbox{\small $\frac{1}{3}$} B_{\mu}) d x_{\mu})
  =  \omega_{SU(3)} \omega_{SU(2)} \omega_{U(1)}^{1/3}, \nonumber\\
 W_{{\rm R}^1_{q}}({\cal C}) & = & \, P \, {\rm exp}( i \int_{\cal C}
 (Z_{\mu} + \mbox{\small $\frac{4}{3}$} B_{\mu}) d x_{\mu})
 = \omega_{SU(3)}\omega_{U(1)}^{4/3}, \nonumber\\
 W_{{\rm R}^2_{q}}({\cal C}) & = & \, P \, {\rm exp}( i \int_{\cal C}
 (Z_{\mu} - \mbox{\small $\frac{2}{3}$} B_{\mu}) d x_{\mu})
 = \omega_{SU(3)}\omega_{U(1)}^{-2/3}, \nonumber\\
 W_H({\cal C}) & = & \, P \, {\rm exp}( i \int_{\cal C}
 ( A_{\mu} +  \ B_{\mu}) d x_{\mu})
 =  \omega_{SU(2)}\omega_{U(1)}, \label{PT}
\end{eqnarray}
where we introduced Wilson loops corresponding to $SU(3)$, $SU(2)$, and $U(1)$
gauge fields respectively:
\begin{eqnarray}
 \omega_{U(1)} & = & {\rm exp}( i \int_{\cal C}  B_{\mu} d x_{\mu}), \nonumber\\
 \omega_{SU(2)} & = &  \, P \, {\rm exp}( i \int_{\cal C}  A_{\mu}
 d x_{\mu}),\nonumber\\
 \omega_{SU(3)} & = &  \, P \, {\rm exp}( i \int_{\cal C}  Z_{\mu} d x_{\mu}).
\end{eqnarray}

In Eq.~(\ref{PT}) each $\omega$ corresponds to a path connecting different
points. However, in  Eq.~(\ref{WWW})  these parallel transporters are arranged
in such a way that  ${\cal W} ({\cal C})$ depends only upon $\omega$'s
corresponding to closed loops constructed of $C_{\alpha}$.

Thus any correlator can be represented through vacuum averages of products of
those loop variables. The vacuum average is considered in the pure gauge theory
with the action $S_g$. Using the loop calculus, we can express $\langle {\cal
W} ({\cal C}) \rangle $ as follows:
\begin{eqnarray}
 && \langle {\cal W} ({\cal C}) \rangle =
 \int D\omega_{U(1)} D\omega_{SU(2)} D\omega_{SU(3)}\times \nonumber\\
 && \times \exp \left( \frac{1}{4} \int \left\{ \frac{1}{3g_{SU(3)}^2} {\rm Tr}\,
 \left.\frac{\delta \omega_{SU(3)}}{\delta \sigma_{\mu \nu}(x)}\right|_0
 \left.\frac{\delta \omega_{SU(3)}}{\delta \sigma_{\mu \nu}(x)} \right|_0
 + \right.\right.  \nonumber \\
 && \left.\left. + \frac{1}{2g_{SU(2)}^2} {\rm Tr}\,
 \left.\frac{\delta \omega_{SU(2)}}{\delta \sigma_{\mu \nu}(x)}\right|_0
 \left.\frac{\delta \omega_{SU(2)}}{\delta \sigma_{\mu \nu}(x)}\right|_0
 \right.\right. + \nonumber\\
 && \left.\left. + \frac{1}{g_{U(1)}^2}
 \left.\frac{\delta \omega_{U(1)}}{\delta \sigma_{\mu \nu}(x)}\right|_0
 \left.\frac{\delta \omega_{U(1)}}{\delta \sigma_{\mu \nu}(x)}\right|_0
 \right\}
 d^4x \right){\cal W} ({\cal C}),
 \label{W1}
\end{eqnarray}
where $\sigma_{\mu \nu}$ is the infinitesimal area, $\frac{\delta}{\delta
\sigma_{\mu \nu}}$ is the area derivative, and $\dots |_0$ means that the area
derivatives are calculated for infinitesimal contours. The measure over the
gauge variables is denoted now as $D \omega$.

\subsection{The symmetry}
Now we are in a position to point out the mentioned symmetry. It turns out that
${\cal W}({\cal C})$, being expressed through $\omega$'s corresponding to
closed loops, is invariant under the following transformation:
\begin{eqnarray}
 \omega_{U(1)}({\cal C}) & \rightarrow & {\rm exp}( -i \pi {\bf L} ({\cal C}, \Sigma))
 \, \omega_{U(1)}({\cal C}), \nonumber\\
 \omega_{SU(2)}({\cal C}) & \rightarrow & {\rm exp}( i \pi {\bf L} ({\cal C}, \Sigma))
 \, \omega_{SU(2)}({\cal C}), \nonumber\\
 \omega_{SU(3)}({\cal C}) & \rightarrow &
 {\rm exp}( i \mbox{\small $\frac{2}{3}$} \pi {\bf L}
 ({\cal C}, \Sigma))\, \omega_{SU(3)}({\cal C}). \label{sym}
\end{eqnarray}
Here, $\Sigma$ is an arbitrary closed surface and ${\bf L} ({\cal C}, \Sigma)$
is the integer linking number of this surface and the closed contour $\cal C$.
From Eq. (\ref{sym}) it is clear that this transformation belongs to $Z_6$
group.

This transformation corresponds to the centers of the $SU(3)$ and $SU(2)$
subgroups of the gauge group. It is finite, being applied to the gauge
invariant loop variables $\omega$.  However, it becomes singular in terms of
gauge potentials:
\begin{eqnarray}
 B_{\mu} & \rightarrow & B_{\mu} -  \pi {\cal V}_{\mu} \nonumber\\
 A_{\mu} & \rightarrow & A_{\mu} +  \pi {\cal V}_{\mu}
 \frac{A_{\nu} t_{\nu} }{({\rm Tr} A_{\tau} t_{\tau})^{1/2}}  \nonumber\\
 Z_{\mu} & \rightarrow & Z_{\mu} +  \frac{2\pi}{3} {\cal V}_{\mu}
 \frac{Z_{\nu} t_{\nu} }{({\rm Tr} Z_{\tau} t_{\tau})^{1/2}},
\label{Transf}
\end{eqnarray}
where ${\cal V}_{\mu}(x) = \int_{V} t_{\mu} \delta(x-y(a,b,c))da \, db\,  dc $
is  an integral over the three-dimensional hypersurface $y(a,b,c)$, the
boundary of which is $\Sigma$. The normal vector to $V$ is denoted by $t_{\mu}
= \frac{1}{2}\epsilon_{\mu \nu \rho \sigma} \frac{\partial y_{\nu}}{\partial a}
\frac{\partial y_{\rho}}{\partial b} \frac{\partial y_{\sigma}}{\partial c}$.

 The invariance of Eq.~(\ref{WWW}) under the transformations
(\ref{sym}) can be easily proven via direct substitution of Eq.~(\ref{Transf})
into Eq.~(\ref{PT}).

\subsection{Redefinition of the gauge action}

The whole SM can be represented (at least, formally) in such a way that it
possesses the symmetry with respect to transformation Eq.~ (\ref{sym}). This
can be done by the following redefinition of the pure gauge part:
\begin{equation}
  \langle {\cal W} ({\cal C}) \rangle =
 \int D\omega_{U(1)} D\omega_{SU(2)} D\omega_{SU(3)}
  \exp \left( \sum_{k} \beta_k \int  {\rm Tr}\,
 \left. \frac{\delta W_{k}}{\delta \sigma_{\mu \nu}(x)} \right|_0
 \left. \frac{\delta W_{k}}{\delta \sigma_{\mu \nu}(x)} \right|_0 \right)
 {\cal W} ({\cal C}).  \nonumber \\
\label {ZW}
\end{equation}
Here the sum is over the six parallel transporters mentioned above. For an
appropriate choice of couplings $\beta_{k}$ the action in Eq.~(\ref{ZW}) is
equal to the action in Eq.~(\ref{W1}) defined in terms of smooth gauge fields.
However, in loop calculus we are not forced to consider smooth gauge fields. We
are allowed to consider {\it piecewise smooth loop variables} $\omega$ instead.
The main difference is that the action in Eq.~(\ref{W1}) suppresses step-like
$\omega$'s, while Eq.~(\ref{ZW}) allows the appearance of loop variables with
Eq.~(\ref{sym})-like discontinuities.

Instead, Eq.~(\ref{ZW}) suppresses the discontinuities in $W_k$. Therefore we
may apply an Eq.~(\ref{sym})-like transformation to all $\omega$'s in order to
make them smooth. After that Eq.~(\ref{ZW}) becomes identical to
Eq.~(\ref{W1}). So, these two formulations would define the same theory.

Here we implied that if the action suppresses some physical quantity, the
latter indeed vanishes. However, there is another point of view. Namely, there
are some indications, that the naively suppressed quantities may survive due to
the entropy factor \cite{BVZ2002}. We do not discuss here this possibility, but
we must mention that if this picture emerges in  the Standard Model,
Eqs.~(\ref{W1}) and (\ref{ZW}) may define different models and correspond to
different physics.  We also notice here that in this case, say, the topological
theta-term with $\theta = 2\pi$ being added to the action could, in principle,
change the nonperturbative behavior of the theory while keeping the same
perturbation expansion \cite{Z2002}.

\subsection{The Standard Model as a finite cutoff theory}

It was mentioned in the introduction that the Standard Model should be regarded
as a finite cutoff theory. So, the correct continuum model must contain a short
distance part (related to the unification of the Electroweak and strong
interactions), which makes the corresponding lattice model cutoff-independent.

The Unified model could be the origin of our additional symmetry. If so,
Eq.~(\ref{sym}) emerges in it without any singular transformation of gauge
potentials. Actually, in the corresponding examples considered in
~\cite{BVZ2003} the realization of Eq.~(\ref{sym}) being written in terms of
the continuum fields is not singular.

Strictly speaking, the only thing we are able to consider is the regularized
model. The regularization can be constructed in such a way that it either
admits or does not admit Eq.~(\ref{sym}). Each choice of regularization is, in
essence, the low energy limit of a regularized Unified model. After the
discretization is removed, the full continuum theory appears. The finite cutoff
Standard Model is an approximation to this hypothetical theory. Our assumption
is that if we construct the {\it finite cutoff Standard Model} (FCSM) either
keeping or not keeping the additional symmetry, the degree to which the
resulting model approximates the correct unified model could be different.

We expect that this difference might manifest itself at high enough energies.
Probably, this could happen in the intermediate region between the usual SM
scale and the GUT scale. Strictly speaking, in this region neither realizations
of the FCSM can describe the physics properly. However, if the symmetry with
respect to Eq.~(\ref{sym}) is indeed a fundamental symmetry, the corresponding
model may give results that are closer to the experimental ones (and vice
versa). If so, we would catch the echo of the Unified model already at
intermediate energies and draw certain conclusions about its structure.

However,  we expect  that the most important role of the symmetry with respect
to (\ref{sym}) is rather technical. The convergence of the lattice methods to
physical results could become considerably faster for the models that respect
invariance under Eq.~(\ref{sym}). This can  be crucial for consideration of
certain processes. Probably, the same situation takes place, say, for the
$SU(2)$ and $SO(3)$ gauge models \cite{SO3}. They are generally believed to
belong to the same universality class. However, physical results are
practically not achievable via $SO(3)$ lattice theory. The reason is that the
$Z_2$ symmetry is lost.

In general, it is thought that the convergence of a lattice model to the
continuum results is faster if it keeps as much symmetries of the continuum
model as possible. It even might occur that models that keep or do not keep a
certain symmetry may lead to different continuum theories. Therefore, we also
do not exclude that FCSM's that respect or do not respect Eq.~(\ref{sym}) would
give essentially different results. In any case nothing definite could be said
until the corresponding numerical research is performed.

\section{The lattice model}

\subsection{Discretization of the continuum model}

In the remaining part of this article we shall not be interested in a
discretization of the fermion sector. We would only notice that there are some
difficulties concerning the problem of keeping the chiral symmetry while
avoiding doubling. There were many different papers on this subject. For a
review see \cite{lattice_fermions} and references therein.

Now our aim is to remind the construction of \cite{BVZ2003}. We construct a
lattice $SU(3)\times SU(2)\times U(1)$ gauge model coupled to the scalar field
in such a way that it reflects all the required properties of the Weinberg -
Salam model and, in addition, preserves the symmetry considered above.

The model contains the following variables:

1. Lattice gauge fields (which live on the links of the lattice):
\begin{eqnarray}
 \Gamma \in SU(3), \quad U \in SU(2), \quad e^{i\theta} \in U(1).
\end{eqnarray}

2. A scalar doublet $\Phi^{\alpha}, \;\alpha = 1,2$ (which lives on the lattice
sites).
The action of the model must have the form:
\begin{equation}
 S = S_g + S_H,
\end{equation}
where we denote by $S_g$ the pure gauge part and the scalar part of the action
is denoted by $S_H$.

A possible choice of $S_H$ is
\begin{equation}
 S_H = \sum_{xy} |U_{xy}e^{-i\theta_{xy}}\Phi_y - \Phi_x|^2
     + \sum_x V(|\Phi_x|),
\end{equation}
where $V(r)$ is the potential, which has a minimum at a nonzero value of $r =
\sqrt{\gamma}$.

To construct the pure gauge part of the action we use the following
correspondence between lattice and continuum notations:
\begin{eqnarray}
 \omega_{U(1)}({\cal C}) & \rightarrow  & \Pi_{{\rm link} \in l}
 e^{-i\theta_{\rm link}},\nonumber\\
 \omega_{SU(2)}({\cal C}) & \rightarrow  & \Pi_{{\rm link} \in l}
 U_{\rm link}, \nonumber\\
 \omega_{SU(3)}({\cal C}) & \rightarrow  & \Pi_{{\rm link} \in l}
 \Gamma_{\rm link},
\end{eqnarray}
where $l$ is a closed contour on the lattice corresponding to the continuum
contour $\cal C$.

The analogue of the continuum transformation is the lattice transformation:
\begin{eqnarray}
 U & \rightarrow & U e^{-i\pi N}, \nonumber\\
 \theta & \rightarrow & \theta +  \pi N, \nonumber\\
 \Gamma & \rightarrow & \Gamma e^{(2\pi i/3)N},
\label{symlat}
\end{eqnarray}
where $N$ is an arbitrary integer link variable.  It represents a
three-dimensional hypersurface on a dual lattice, the boundary of which
corresponds to $\Sigma$ in Eq.~(\ref{sym}). This symmetry reveals the
correspondence between the centers of the $SU(2)$ and $SU(3)$ subgroups of the
gauge group.

The choice $\beta = \beta_{{\rm L}_{\ell}} = \beta_{{\rm R}_{\ell}} =
\beta_{{\rm L}_{q}} = \beta_{{\rm R}^1_{q}} = \beta_{{\rm R}^2_{q}}$ and
$\beta_H = 0 $ corresponds to a certain class of unified models \cite{BVZ2003}.
Hence, we choose:
\begin{eqnarray}
 S_g & = & \beta \sum_{\rm plaquettes}
 (2(1-\mbox{${\small \frac{1}{2}}$} {\rm Tr}\, U_p \cos \theta_p)+
 \nonumber \\
 && +(1-\cos 2\theta_p) +\nonumber \\
 && +6(1-\mbox{${\small \frac{1}{6}}$} {\rm Re Tr}
 \,\Gamma_p {\rm Tr}\, U_p {\rm exp} (i\theta_p/3))+
 \nonumber\\
 && +3(1-\mbox{${\small \frac{1}{3}}$} {\rm Re Tr}
 \, \Gamma_p {\rm exp} (-2i\theta_p/3)) +\nonumber \\
 && +3(1-\mbox{${\small \frac{1}{3}}$} {\rm Re Tr}
 \, \Gamma_p {\rm exp} (4i\theta_p/3))),\label{Act}
\end{eqnarray}
where the sum runs over the elementary plaquettes of the lattice. Each term of
the action Eq.~(\ref{Act}) corresponds to a parallel transporter along the
boundary  $\partial p$ of plaquette $p$. The correspondent plaquette variables
constructed of lattice gauge fields are $U_p = \omega_{SU(2)}(\partial p)\, ,
\Gamma_p = \omega_{SU(3)}(\partial p)\, $, and $ \theta_p = {\rm Arg}\,
\omega_{U(1)}(\partial p)$.

\subsection{The simplified model}
In this paper we report our results on the numerical simulation of the model,
in which we omit the dynamical fermions, as well as the color subgroup $SU(3)$.
It will be seen below that already on this level certain qualitative
differences between this model and the conventional one exist.

The potential for the scalar field is considered in the London limit, i.e., in
the limit of infinite bare Higgs mass. The action of the model reduces to
\begin{eqnarray}
 S & = & S^{\rm L} + S^{\rm R} + S_H =
 \nonumber\\
 & = & \beta \!\! \sum_{\rm plaquettes} \!\!
 ((1- \mbox{${\small \frac{1}{2}}$} \,{\rm Tr}\, U_p \cos \theta_p)
 + \mbox{${\small \frac{1}{2}}$} (1-\cos 2\theta_p)) +\nonumber\\
 && +\sum_{xy} |U_{xy}e^{-i\theta_{xy}}\Phi_y - \Phi_x|^2 +
 V(|\Phi|).\label{S_num}
\end{eqnarray}
(Here $\beta$ is rescaled as $\beta \rightarrow \beta/2$ for the convenience of
comparing the results with those of the $SU(2)$ fundamental Higgs model.)
$S^{\rm L}$ corresponds to the  doublet of left-handed fermions and $S^{\rm R}$
corresponds to the right-handed singlet.  $\Phi$ is the Higgs doublet and $V$
is an infinitely deep potential, giving rise to the vacuum average $\langle
|\Phi| \rangle  = \sqrt{\gamma}$.  It is worth mentioning that the naive
continuum limit of Eq.~(\ref{S_num}) gives the value of Weinberg angle
$\theta_W = \pi/6$, which is surprisingly close to the experimental value.

After fixing the unitary gauge we obtain:
\begin{eqnarray}
 S & = & \beta \!\! \sum_{\rm plaquettes}\!\!
 ((1-\mbox{${\small \frac{1}{2}}$} \, {\rm Tr}\, U_p \cos \theta_p)
 + \mbox{${\small \frac{1}{2}}$} (1-\cos 2\theta_p)+\nonumber\\
 && + \gamma \sum_{xy}(1 - Re(U^{11}_{xy} e^{-i\theta_{xy}}))).
\end{eqnarray}
Of course, we keep in mind that this simplification of the model may lead to
some qualitative changes in the description of the dynamics. Thus the
conclusions, which we draw after performing the numerical investigation of the
simplified model, must be justified by the study of the full model, including
the color subgroup, dynamical fermions, and a finite Higgs mass.

Below we briefly describe some of the quantities which we investigate in this
work.

The following variables are considered as creating a photon, $Z$ boson, and $W$
boson respectively:
\begin{eqnarray}
 A_{xy} & = & A^{\mu}_{x} \; = \,[{\rm Arg} U_{xy}^{11} + \theta_{xy}]
 \,{\rm mod} \,2\pi, \nonumber\\
 Z_{xy} & = & Z^{\mu}_{x} \; = \,[{\rm Arg} U_{xy}^{11} - \theta_{xy}]
 \,{\rm mod} \,2\pi, \nonumber\\
 W_{xy} & = & W^{\mu}_{x} \,= \,U_{xy}^{12} e^{i\theta_{xy}}.
\end{eqnarray}
Here, $\mu$ represents the direction $(xy)$. After fixing the unitary gauge the
electromagnetic $U(1)$ symmetry remains:
\begin{eqnarray}
 U_{xy} & \rightarrow & g^\dag_x U_{xy} g_y, \nonumber\\
 \theta_{xy} & \rightarrow & \theta_{xy} +  \alpha_y/2 - \alpha_x/2,
\end{eqnarray}
where $g_x = {\rm diag} (e^{i\alpha_x/2},e^{-i\alpha_x/2})$. The fields $A$,
$Z$, and $W$ transform as follows:
\begin{eqnarray}
 A_{xy} & \rightarrow & A_{xy} + \alpha_y - \alpha_x, \nonumber\\
 Z_{xy} & \rightarrow & Z_{xy}, \nonumber\\
 W_{xy} & \rightarrow & W_{xy}e^{-i\alpha_x}.
\label{T}
\end{eqnarray}

As any other compact gauge theory, our model contains monopoles. As in other
compact gauge models, their behavior is connected with the possible confinement
of charges. On the other hand, the continuum Weinberg Salam model is believed
not to confine any charges and not to be affected by monopoles.

We investigated two types of monopoles. $U(1)$ monopoles extracted from
$2\theta$ are defined as
\begin{equation}
 j_{2\theta} = \frac{1}{2\pi} {}^*d([d 2\theta]{\rm mod}2\pi).
\end{equation}
The electromagnetic monopoles are:
\begin{equation}
 j_{A} = \frac{1}{2\pi} {}^*d([d A]{\rm mod}2\pi) .
\end{equation}
(Here we used the notations of differential forms on the lattice. For the
definition of those notations see, for example, ~\cite{forms}.)

The density of the monopoles is defined as follows:
\begin{equation}
 \rho = \left\langle \frac{\sum_{\rm links}|j_{\rm link}|}{4L^4} \right\rangle,
\end{equation}
where $L$ is the lattice size.
To understand the dynamics of external charged particles, we consider the
Wilson loops defined in the representations of left-handed and right-handed
leptons:
\begin{eqnarray}
 {\cal W}^{\rm L}(l) & = & \langle {\rm Re} {\rm Tr} \,\Pi_{(xy) \in l}
 U_{xy} e^{-i\theta_{xy}}\rangle, \nonumber\\
 {\cal W}^{\rm R}(l) & = & \langle {\rm Re} \Pi_{(xy) \in l}
 \, e^{-2i\theta_{xy}}\rangle .
\end{eqnarray}
Here $l$ denotes a closed contour on the lattice. We consider the following
quantity constructed from the rectangular Wilson loop of size $a\times a$:
\begin{equation}
{\cal V}_{R,L}(a) = - \log {\cal W}^{R,L}(a\times a)/a.
\end{equation}
A linear behavior of ${\cal V}(a)$ would indicate the existence of a charge -
anti charge string with nonzero tension.

\subsection{Numerical results}

In our calculations we investigated lattices $L^4$ for $L = 6$, $L = 12$, and
$L = 16$ with symmetric boundary conditions.

We summarize our qualitative results in the phase diagram represented in
Fig.~\ref{fig.1}. The model contains three phases.  The first one (I) is a
confinement-like phase, in which the dynamics of external charged particles is
similar to that of QCD with dynamical fermions. In the second phase (II) only
the behavior of left-handed particles is confinement-like, while for
right-handed ones it is not. The last one (III) is the Higgs phase, in which no
confining forces are observed at all.  This is illustrated by Figs.~\ref{fig.5}
and \ref{fig.6}, in which we represent ${\cal V}_L(a)$ and ${\cal V}_R(a)$ at
three typical points that belong to different phases of the model. One can see
that in the Higgs phase the shape of ${\cal V}(a)$ excludes the possibility of
a linear potential to exist. The same behavior is found in phase II for ${\cal
V}_R(a)$. On the other hand, in phase II the shape of ${\cal V}_L(a)$ signals
the appearance of a linear potential at sufficiently small distances (up to
five lattice units).  However, as for QCD with dynamical fermions or the
$SU(2)$ fundamental Higgs model \cite{Montvay,EW_T}, these results do not mean
that confinement occurs.  The charge - anti charge string must be torn by
virtual charged scalar particles, which are present in the vacuum due to the
Higgs field. Thus ${\cal V}(a)$ may be linear only at sufficiently small
distances, while starting from some distance it must not increase, indicating
the breaking of the string.  Unfortunately the accuracy of our measurements
does not allow us to observe this phenomenon in detail.  However, it may be
partially illustrated by the shapes of ${\cal V}_L(a)$ and ${\cal V}_R(a)$ in
phase I shown in Fig.~\ref{fig.5} and Fig.~\ref{fig.6}.

The phase structure of the model may also be seen through the data for the mean
action over the whole lattice $\bar{S} = \langle S \rangle /(6\beta L^4)$,
Fig.~\ref{fig.2}. It appears to be inhomogeneous in a small vicinity of the
phase transition line.

The connection between the properties of monopoles and the phase structure of
the model is illustrated by Figs.~\ref{fig.3} and \ref{fig.4}, which show the
monopole density versus the coupling constants. The electromagnetic monopole
density drops in the Higgs phase, while the $U(1)$ monopole density falls
sharply both in phase II and phase III. We can see that the behavior of the
$U(1)$ monopoles is connected with the dynamics of the right-handed particles,
while the behavior of electromagnetic monopoles reflects the dynamics of the
left-handed particles.

It is worth mentioning that the cousin of our model, the $SU(2)$ fundamental
Higgs model,  has a similar phase structure as our model, except for the
absence of the phase transition line between phases I and II. In the latter
model it was shown that different phases are actually not different. This means
that the phase transition line ends at some point and the transition between
two states of the model becomes continuous. Thus one may expect that in our
model the phase transition line between phases I and III ends at some point.
However, we do not observe this  for the considered values of couplings.

In our model both phase transition lines join in a triple point, forming the
common line. This is, evidently, the consequence of the mentioned additional
symmetry that relates $SU(2)$ and $U(1)$ excitations. The same picture, of
course,  does not emerge in the conventional $SU(2) \otimes U(1)$ gauge --
Higgs model \cite{SU2U1}.

We must also notice here that the phase diagram may also contain an unphysical
region, corresponding to the unphysical region of the pure $SU(2)$ model (which
is observed at $\beta < \beta_c$, where $\beta_c$ is the crossover point). Our
investigation shows that if this region of couplings exists in our model, it
must be far from the Higgs phase, which is of main interest to us. Indeed, this
unphysical region might appear for $\beta < 2.25$ and $\gamma < 0.5$.

\section{Conclusions}

We summarize our results as follows:

1. We illustrated an additional symmetry found in the fermion and the Higgs
sectors of the Standard Model by the consideration of the SM in loop space.

2. We performed a numerical investigation of the quenched Electroweak sector of
the lattice model, that respects the additional symmetry.

3. The lattice model contains three phases. The first one is a confinement-like
phase.  In the second phase the confining forces are observed, at sufficiently
small distances, only between the left-handed particles.  The last one is the
Higgs phase.

4. The main consequence of the emergence of the additional symmetry is that the
phase transition lines corresponding to the $SU(2)$ and $U(1)$ degrees of
freedom join in a triple point forming the common line. This reflects the fact
that the $SU(2)$ and $U(1)$ excitations are related due to the mentioned
symmetry. The same situation does not take place in the conventional
$SU(2)\otimes U(1)$ gauge - Higgs model \cite{SU2U1}.

So, already on this simplified level, we found a qualitative difference between
the conventional discretization and the discretization that respects the
invariance under Eq.~(\ref{sym}).

\begin{acknowledgments}
We are grateful to M.I. Polikarpov and F.V. Gubarev for useful discussions.
A.I.V. and M.A.Z. kindly acknowledge the hospitality of the Department of
Physics and Astronomy of the Vrije Universiteit, where part of this work was
done.  We also appreciate R. Shrock and I.Gogoladze, who have brought to our
attention references \cite{SU2U1} and \cite{Z6} respectively. This work was
partly supported by RFBR grants
 03-02-16941, 04-02-16079 and 02-02-17308, by the INTAS grant 00-00111, the CRDF
award RP1-2364-MO-02, DFG grant 436 RUS 113/739/0, and RFBR-DFG grant
03-02-04016, by Federal Program of the Russian Ministry of Industry, Science,
and Technology No 40.052.1.1.1112.  \end{acknowledgments}

\newpage

\begin{figure}
\begin{center}
 \epsfig{figure=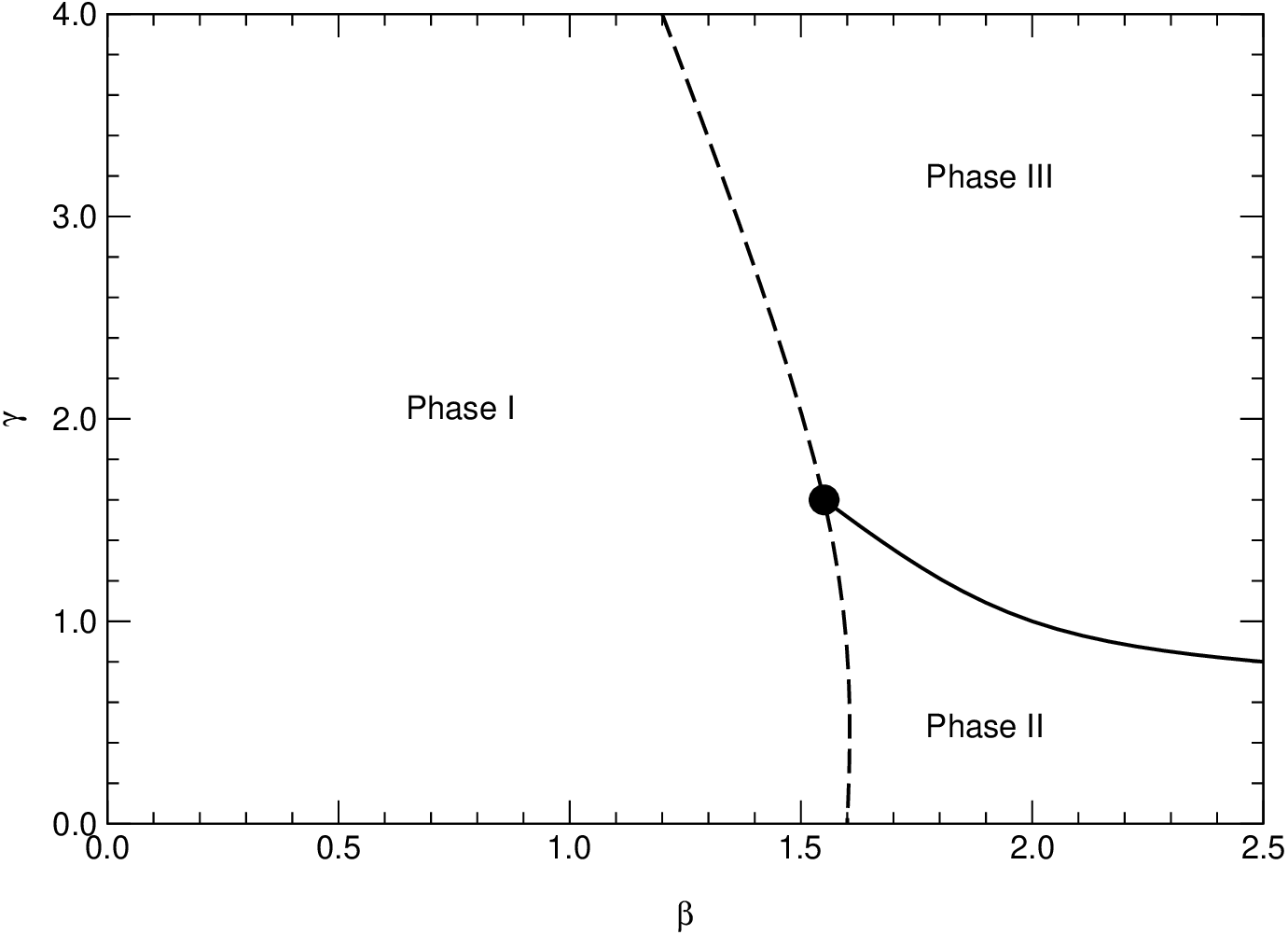,height=60mm,width=80mm,angle=0}
 \caption{\label{fig.1} The phase diagram of the model in the
 $(\beta, \gamma)-plane$.}
\end{center}
\end{figure}

\begin{figure}
\begin{center}
 \epsfig{figure=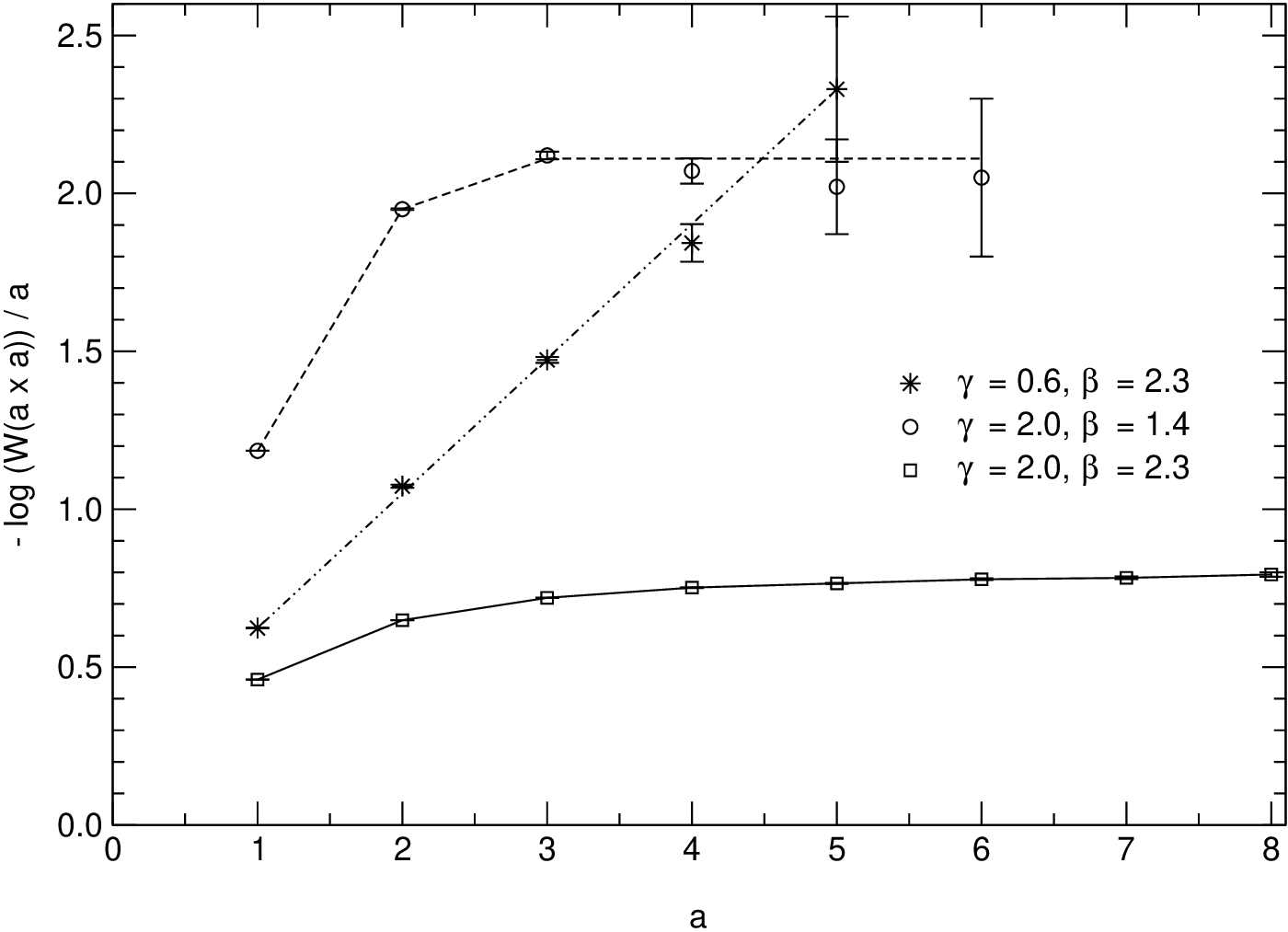,height=60mm,width=80mm,angle=0}
 \caption{\label{fig.5} ${\cal V}_L(a)$ calculated
  at three points that belong to different phases
  of the model.}
\end{center}
\end{figure}

\begin{figure}
\begin{center}
 \epsfig{figure=Wcu.eps,height=60mm,width=80mm,angle=0}
 \caption{\label{fig.6}  ${\cal V}_R(a)$ calculated
  at three points that belong to different phases
  of the model.}
\end{center}
\end{figure}

\begin{figure}
\begin{center}
 \epsfig{figure=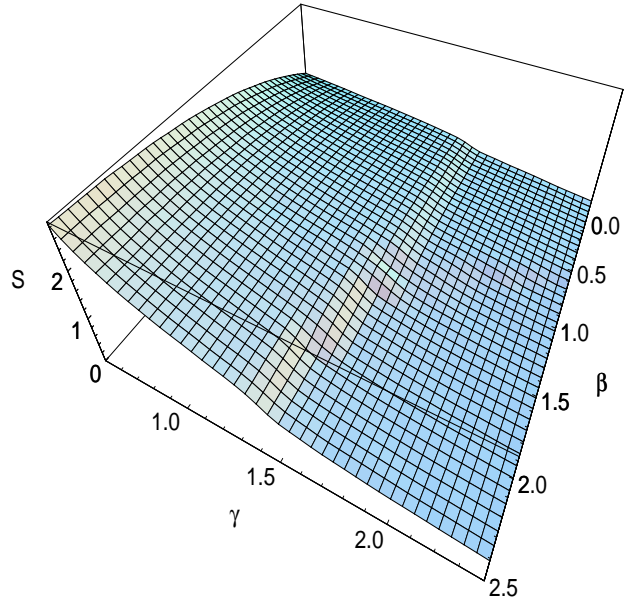,height=80mm,width=80mm,angle=0}
 \caption{\label{fig.2} The action
 $\bar{S} =  \langle S \rangle /(6\beta L^4)$ of the model.}
\end{center}
\end{figure}

\begin{figure}
\begin{center}
 \epsfig{figure=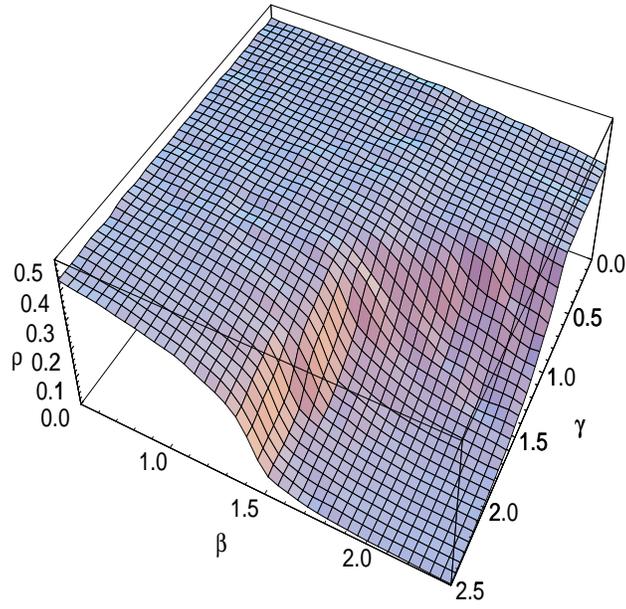,height=80mm,width=80mm,angle=0}
 \caption{\label{fig.3} The density of electromagnetic monopoles.
 It decreases in the Higgs phase.}
\end{center}
\end{figure}

\begin{figure}
\begin{center}
 \epsfig{figure=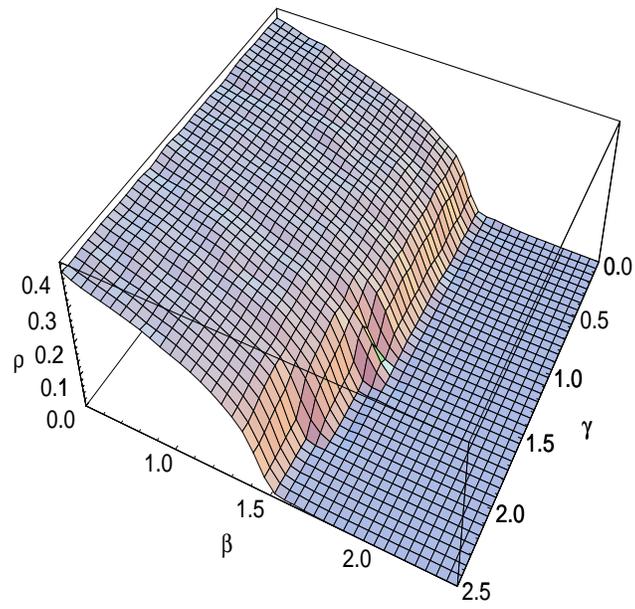,height=80mm,width=80mm,angle=0}
 \caption{\label{fig.4} The density of $U(1)$ - monopoles. It decreases
when the behavior of right - handed external particles is not
confinement - like.}
\end{center}
\end{figure}

\end{document}